\documentclass[sigconf]{acmart}

\usepackage[utf8]{inputenc}
\usepackage[T1]{fontenc}
\usepackage{graphicx}
\usepackage{enumitem}
\usepackage{booktabs}

\copyrightyear{2026}
\acmYear{2026}
\setcopyright{cc}
\setcctype{by}
\acmConference[IDC '26]{Proceedings of the 25th Interaction Design and Children Conference}{June 22--25, 2026}{Brighton, United Kingdom}
\acmBooktitle{Proceedings of the 25th Interaction Design and Children Conference (IDC '26), June 22--25, 2026, Brighton, United Kingdom}
\acmDOI{10.1145/3773077.3812190}
\acmISBN{979-8-4007-2283-7/2026/06}

\begin{document}

\title[Social Understanding, Placeness, and Identity Alignment:\\A Design Framework for Friendship-Supportive Youth Social Media]{Social Understanding, Placeness, and Identity Alignment: A Design Framework for Friendship-Supportive Youth Social Media}

\author{JaeWon Kim}
\affiliation{%
  \institution{University of Washington}
  \city{Seattle}
  \state{WA}
  \country{USA}}
\email{jaewonk@uw.edu}

\author{Alexis Hiniker}
\affiliation{%
  \institution{University of Washington}
  \city{Seattle}
  \state{WA}
  \country{USA}}
\email{alexisr@uw.edu}

\begin{abstract}
We present a design framework for friendship-supportive youth social media, derived from a synthesis of five empirical studies with 331 youth participants (ages 13--25) using interviews, co-design, surveys, diary studies, and a field deployment. Iterative analysis of 209 design-relevant data points identified three pillars: \textit{Social Understanding} (interaction norms, interaction cues and scaffolding, social accountability and governance), \textit{Placeness} (third place and community, boundaries and personal spaces, shared presence), and \textit{Identity Alignment} (identity currency, identity plurality, relational identity signals). The framework maps nine design spaces through which platforms can support the conditions under which youth friendships form, deepen, and are maintained. It offers a shared vocabulary for locating contributions, comparing design interventions, and identifying under-explored areas for future work.
\end{abstract}



\maketitle

\section{Introduction}

Youth friendships depend on legible social cues, stable environments where familiarity can develop through repeated contact, and room for identity to evolve as adolescents grow~\cite{erikson1994, altman1973, oldenburg1999}. Many current social media platforms provide these conditions unevenly. Systems optimized for engagement metrics, content virality, and broad audience reach often reward curation and raise the social cost of vulnerability, even as youth report wanting more authentic connection~\cite{yau2019, barta2021}.

Researchers across communication studies, developmental psychology, environmental psychology, and social computing have diagnosed pieces of this problem. Electronic media can erode the sense of place that structures social life~\cite{meyrowitz1986no}; persistent profiles and collapsed audiences can conflict with adolescents' need for fluid self-presentation~\cite{erikson1994, marwick2011}; and platform affordances shape the norms that govern self-disclosure and trust~\cite{erickson2000, masur2023}. These literatures explain why norms, place, trust, and identity matter, but they do not by themselves show how to translate those concerns into concrete design choices. The gap is one of design translation: naming the distinct areas of the design space where platforms can support the formation, deepening, and maintenance of youth friendships.

We define \textit{friendship-supportive design} as design that creates conditions for recurring relational processes, including low-risk initiation, reciprocal self-disclosure, trust calibration, shared experience, and identity recognition. We present a framework organized around three pillars: \textit{Social Understanding}, \textit{Placeness}, and \textit{Identity Alignment}. The framework emerges from a cross-study synthesis of five empirical studies using interviews, co-design, surveys, diary methods, and a crossover field deployment with 331 youth participants aged 13--25. Through iterative thematic analysis of 209 design-relevant data points, we mapped recurring mechanisms, grouped them by relational function, and refined a three-pillar, nine-dimension structure.

We offer the framework as a map of design spaces rather than a causal model. A single feature may touch multiple dimensions, and design choices may create tensions across them, but the dimensions remain analytically separable because each names a different kind of support youth social media may need to provide. Our contribution is to operationalize established theoretical concerns, including social penetration, social translucence, context collapse, and no sense of place, as designable affordances, defaults, interaction patterns, and feature families for youth social media.

\section{Related Work}

\subsection{Social Norms and Interpersonal Legibility Online}

Social norms shape behavior through informational cues, anticipated consequences, and shortcuts under ambiguity~\cite{morris2015}. Descriptive, injunctive, and subjective norms independently predict self-disclosure on social media~\cite{masur2023}. Yet adolescents often encounter platforms where the visible norm is polished self-presentation, even when many privately want more authenticity~\cite{yau2019, zillich2021, waterloo2018}. This creates social risk: vulnerability can feel deviant because youth cannot easily tell what others actually value. Conversely, platform norms can make privacy-invasive behaviors feel socially acceptable when they are framed as communal or prosocial~\cite{Gilbert2025-ib}.

Design shapes these dynamics. Interface elements foreground what others do, while prompts, defaults, and feedback mechanisms communicate what kinds of behavior are expected~\cite{erickson2000}. Social translucence argues that making social information visible can help people coordinate shared expectations~\cite{erickson2000}. At the same time, online interaction strips away many offline cues, including body language, tone, timing, and proximity, that help people judge intent and receptiveness~\cite{walther1992, ye2025navigating}. Friendship-supportive platforms must therefore make interaction norms and cues legible enough for youth to navigate disclosure, reciprocation, and trust across different levels of intimacy~\cite{altman1973}.

\subsection{Spatial Contexts and the Absence of Place Online}

Relationships develop through repeated, situated contact. Physical third places such as cafes, parks, and community centers support tie formation through spatial propinquity, stable return visits, and boundaries that separate public from private interaction~\cite{oldenburg1999, small2019}. Meyrowitz described electronic media as producing a ``no sense of place,'' weakening the link between setting and social role~\cite{meyrowitz1986no}. Dourish similarly distinguishes abstract \textit{space} from socially meaningful \textit{place}, arguing that design must attend to how environments become lived contexts through use~\cite{dourish2006}.

Digital environments can also become places. People reason about online interaction through spatial metaphors such as ``rooms,'' ``neighborhoods,'' and ``architecture,'' and these metaphors draw on embodied ways of navigating social proximity and distance~\cite{wilson2002six, hall1968proxemics}. Most social media, however, operates as placeless streams: feeds move past without stable gathering spots, clear boundaries, or ambient awareness of who is around. As physical third places decline~\cite{rao2024}, youth social media needs guidance for designing digital places rather than content flows alone.

\subsection{Identity Development and Self-Presentation in Digital Contexts}

Adolescence involves rapid identity change, heightened peer sensitivity, and the need to try on different versions of the self~\cite{erikson1994, steinberg2005}. Online self-presentation is complicated by context collapse, where diverse audiences flatten into a single public~\cite{marwick2011}, and by persistence, which can attach youth to outdated expressions long after they no longer feel accurate. Goffman's dramaturgical account of impression management highlights why forcing one stable presentation of self conflicts with ordinary social life~\cite{goffman2016}.

Feature-level responses such as ephemerality, audience segmentation, and pseudonymity address parts of this tension~\cite{bayer2016}. But they are often treated as isolated mechanisms rather than as components of a broader design space. Supporting identity on youth social media requires asking whether young people can keep self-representations current, express multiple facets across contexts, and provide enough early identity signals for potential friends to approach them appropriately~\cite{walther2002cues}.

\subsection{Friendship-Supportive Design}
Recent work argues that designing social technology for flourishing requires moving beyond paradigms of preventing harm to actively cultivating positive social experience~\cite{kim2024positech, kim2025positech}. Friendship-supportive design does not seek to produce friendships directly. Rather, it improves the conditions under which friendships can form, deepen, and be maintained. Friendship develops through recurring processes such as initiation, reciprocal self-disclosure, calibration of trust and intimacy, repeated shared experience, and recognition of one another's changing identities~\cite{altman1973, erikson1994}. We treat these as cross-cutting relational goals. The framework's nine dimensions do not map one-to-one onto formation, deepening, or maintenance; instead, each names an environmental condition that can support those processes differently across friendship stages~\cite{conwill2025design, alluhidan2024teen}.

\section{Method}

The framework derives from a cross-study synthesis of five empirical studies conducted as part of a larger research program on youth social media design. Together, the studies examined how platform design shapes youth social experience across semi-structured interviews, co-design sessions, a survey, diary studies, and a crossover field deployment, with 331 youth participants in total (ages 13--25).

\textbf{Study A}~\cite{kim2024bereal} interviewed 29 teens (ages 13--18) about an ephemeral photo-sharing platform, showing how platform-enforced authenticity can both lower performance pressure and constrain deliberate self-expression. \textbf{Study B}~\cite{kim2025privacy, kim2025trust} combined co-design interviews with 19 teens, a one-week diary study, and a survey with 136 teens (ages 13--18) to examine privacy fears, trust-based disclosure, and privacy-supportive design. \textbf{Study C}~\cite{kim2025discord} interviewed 25 users (ages 15--24) of a community-oriented platform, identifying how spatial persistence, governance structures, and tiered engagement foster third-place experiences. \textbf{Study D}~\cite{kim2026hogwarts} used Fictional Inquiry co-design with 23 youth (ages 15--24), revealing that youth often imagine social media as navigable spatial environments rather than content feeds. \textbf{Study E} deployed a custom social media platform in a four-week crossover study with 99 youth (ages 15--25) across two countries, comparing relationship-scaffolding features against a conventional baseline through behavioral logs and interviews.

To synthesize findings, we extracted design principles, features, and theoretical takeaways across the five studies, yielding 209 design-relevant data points. We inductively coded these points by conceptual similarity and relational function, then refined the structure through four rounds of coding. This version of the framework is a work in progress: its three-pillar structure is stabilizing, but future studies may refine, extend, or reorganize specific dimensions.

\section{Framework: Three Pillars of Friendship-Supportive Design}

The framework is organized around three pillars: \textit{Social Understanding}, \textit{Placeness}, and \textit{Identity Alignment}. We describe each pillar and its sub-dimensions below.

\subsection{Pillar 1: Social Understanding}

This pillar captures the affordances through which platforms make the social situation legible: what behavior is normal, what an action is likely to mean, when initiation feels welcome, and how trust can be calibrated.

\subsubsection{Interaction Norms.}
Interaction norms, including descriptive and injunctive norms, provide behavioral templates that reduce uncertainty in ambiguous social situations. Platform design can establish these norms through defaults, prompts, interaction templates, and visibility patterns. In Study E, daily question prompts reframed posting as communal participation rather than individual performance. A profile-list feed that gave every user equal visual space, regardless of posting frequency, also reduced anxiety about ``taking up space'' and nearly doubled posting volume. These findings suggest that platforms can make low-stakes participation feel ordinary when their interfaces visibly distribute attention and normalize everyday sharing~\cite{popowski2024commit}.

\subsubsection{Interaction Cues and Scaffolding.}
Interaction cues and scaffolds reduce ambiguity around initiating, reciprocating, and escalating interaction. Features such as prompts, directed questions, and lightweight reactions can compensate for missing offline cues by signaling intent while lowering social risk. In Study E, directed prompt-sending allowed a user to send a specific question to a particular friend, communicating attention and care while giving the recipient a ready-made reason to respond. Self-described lurkers reported posting more when these scaffolds reduced the effort and vulnerability of participation. Different relationship stages require different scaffolds: weaker ties benefit from low-commitment openings and plausible deniability, while close ties can sustain more direct signals of care.

\subsubsection{Social Accountability and Governance.}
Social accountability and governance refer to mechanisms that make actions interpretable, traceable, and socially consequential while allowing communities to articulate and enforce expectations over time. In Study C, collaboratively maintained guidelines made community norms visible and negotiable rather than implicit~\cite{fiesler2018reddit, dym2020social}. In Study B, teens proposed a ``red flag and follow-up'' feature in which users could report boundary violations and receive confirmation when action was taken~\cite{kim2025privacy}. This dimension treats accountability not only as moderation, but also as social infrastructure that helps users understand what happened, what responses are available, and whether community expectations are being maintained.

\subsection{Pillar 2: Placeness}

This pillar captures the spatial, embodied, and environmental affordances that let a platform feel inhabitable: how users sense where they are, who else is present, and what forms of boundary, ambience, and co-presence are possible. Third-place experience is one branch of this pillar, not the pillar as a whole~\cite{meyrowitz1986no, leonardi2015ambient}.

\subsubsection{Third Place and Community.}
Third places are digital gathering places people can return to, where shared history, inside jokes, regulars, and implicit expectations accumulate. In Study C, persistent themed channels functioned as gathering spots where community culture developed through repeated presence~\cite{kim2025discord}. Recurring activities, such as game sessions and daily check-ins facilitated by bots, gave users low-pressure reasons to show up and interact. Persistent, segmented spaces and recurring communal rituals therefore help users recognize regulars, build shared references, and gradually integrate into communities through repeated low-stakes contact.

\subsubsection{Boundaries and Personal Spaces.}
Boundaries shape how widely content travels and how exposed or contained users feel. In Study D, youth drew on spatial metaphors from physical life: bedrooms signaled intimate spaces, common rooms signaled shared spaces, and neighborhoods signaled broader communities~\cite{kim2026hogwarts}. These metaphors made audience scope and social expectations easier to reason about. Across studies, teens worried about content leaking between close friends and broader audiences, and about unclear lines around what counts as private~\cite{kim2025trust}. Friendship-supportive platforms therefore need boundaries that are not only configurable, but also intuitively legible~\cite{zhang2025burst}.

\subsubsection{Shared Presence.}
Shared presence is the felt sense of being around others in a digital environment, whether or not direct interaction occurs. Lightweight ambient awareness can support familiarity without demanding response. Across studies, youth wanted presence signals such as activity indicators, currently-listening displays, and emoji-based statuses that communicate availability or mood without forcing conversation~\cite{kim2025discord}. In Study D, participants imagined sitting together in a virtual room or walking alongside someone asynchronously~\cite{kim2026hogwarts}. These examples suggest that presence is not simply synchronous interaction; it is the background awareness that lets a space feel inhabited.

\subsection{Pillar 3: Identity Alignment}

This pillar captures the fit between an evolving self, the expressive surfaces a platform provides, and the social interpretations those expressions invite.

\subsubsection{Identity Currency.}
Identity currency refers to how easily youth can keep outward identity signals up to date with their current sense of self. Features such as ephemerality, soft deletion, natural decay, and audience recalibration reduce the persistence of outdated self-presentations. In Study E, users could retroactively apply new privacy settings when changing a friendship designation, preventing earlier self-presentations from being read as current identity claims. In Study B, periodic prompts to revisit old posts helped teens reassess whether earlier content still felt accurate without requiring constant vigilance~\cite{kim2025privacy}.

\subsubsection{Identity Plurality.}
Identity plurality refers to supporting multiple coexisting facets of self-expression across contexts, audiences, and expressive modalities. In Study E, varied question prompts invited playful, serious, and reflective responses, allowing users to show different dimensions of themselves without being reduced to one persona. In Study D, youth imagined houses they could own and decorate, where room design and artifact placement would communicate personality indirectly while leaving room for interpretation~\cite{kim2026hogwarts}. This dimension foregrounds identity as multiple and situated rather than singular and fixed.

\subsubsection{Relational Identity Signals.}
Relational identity signals are lightweight cues that help people form an initial sense of one another before deciding whether and how to connect. These cues include shared interests, mutual affiliations, roles, and visible traces of participation. In Study C, shared server memberships and overlapping activity contexts helped users infer potential compatibility~\cite{kim2025discord}. Observing how someone participated across channels gave users a sense of tone, interests, and interaction style, making approach feel less risky. This dimension is especially important for new or casual connections, where the absence of context can make initiation difficult.

\section{Discussion}

\subsection{Contribution and Use}

The framework's main contribution is a design vocabulary for friendship-supportive youth social media. Rather than proposing a causal model, it translates concerns from social penetration, social translucence, no sense of place, context collapse, and adolescent identity development into feature-level questions: Are norms legible? Are spaces persistent and bounded? Are interaction cues interpretable? Can identity signals stay current, plural, and relationally meaningful? The resulting map helps researchers locate what an intervention supports, compare platforms at the level of design spaces rather than individual features, and identify areas where empirical design knowledge remains thin. Some dimensions, such as interaction norms and audience boundaries, already have substantial grounding in social computing; others, including identity currency, shared presence, and relational identity signals, need more targeted study as youth social media design problems.

\subsection{Tensions Across Dimensions}

Dimensions often overlap, but their tensions make them useful to separate. Accountability can make behavior interpretable and repairable while constraining identity plurality if every contribution becomes too visible, persistent, or traceable. Audience segmentation can support safer disclosure but fragment the shared presence that makes a community feel inhabited. Directed prompts can signal care and support deeper ties, but they may also remove the plausible deniability that makes lightweight initiation safer for weaker ties. Naming these tradeoffs helps researchers specify which design space a system advances, which it leaves untouched, and which adjacent concerns it may strain~\cite{akter2025calculating}.

\subsection{Mapping Existing Platforms}

The framework can also describe existing systems. Discord illustrates dense support for place through persistent themed channels, recognizable regulars, stable communal structure, roles, rules channels, and moderation systems~\cite{kim2025discord}. At the same time, persistent message histories may weaken identity currency by attaching users to past expressions after their interests, relationships, or self-understandings have shifted. BeReal illustrates dense support for interaction norms through daily prompts and reciprocal posting requirements that make casual sharing feel expected rather than attention-seeking~\cite{kim2024bereal, li2025bereal}. Its synchronized posting window also creates a weak shared rhythm, but its single-prompt format leaves little room for identity plurality. These contrasts show how platforms can support one design space densely while leaving another underspecified.

\subsection{Scope and Future Work}

The claims are strongest for youth-facing systems involving sharing, profiles, audience management, and peer interaction. The synthesis draws on five studies with participants ages 13--25, mostly in the United States, with one field deployment spanning the United States and South Korea. Messaging-first systems, younger children, other cultural contexts, and platforms with different governance or economic models may require revisions to the framework. Future work should test, refine, and extend each dimension, especially where empirical design knowledge remains thin, and work with practitioners to turn the framework into actionable design guidance.

\section{Conclusion}
The three-pillar framework maps nine design spaces for friendship-supportive youth social media. Rather than treating social understanding, placeness, and identity alignment as abstract ideals, it identifies concrete areas where platform design can shape the conditions under which youth friendships form, deepen, and are maintained: legible social dynamics, stable and inhabitable places, and identity expression that remains current, plural, and relationally interpretable. Grounded in a synthesis of five empirical studies with 331 youth participants, the framework offers a shared vocabulary for situating new work, identifying under-explored areas, and designing platforms where youth friendships can develop. In doing so, it shifts the question from whether youth social media is good or bad for young people to what it should be built to do.

\section{Selection and Participation of Children}
Participants across the five studies were aged 13--25 and recruited through social media advertisements, community outreach, and snowball sampling. For minors, we obtained informed parental or guardian consent alongside participant assent prior to any study activities. All protocols were reviewed and approved by our university's Institutional Review Board, and participants received age-appropriate compensation. Participants were informed of their right to withdraw at any time. In the field deployment, minors used the platform under the same consent protocols, with content monitoring in place throughout the study period.

\begin{acks}
JaeWon Kim would like to acknowledge the CERES Network, University of Washington Global Innovation Funds (GIF), and Student Technology Funds (STF), which provided support for this work. This work was also funded in part by the Paul G. Allen School of Computer Science \& Engineering Endowed Fund for Excellence and a gift from Google. Alexis Hiniker is a special government employee for the Federal Trade Commission. The content expressed in this manuscript does not reflect the views of the Commission or any of the Commissioners.
\end{acks}

\bibliographystyle{ACM-Reference-Format}
\bibliography{references,references1,references2,references3,references_supp}

\end{document}